\documentstyle[sprocl,psfig]{article}

\def\Journal#1#2#3#4{{#1} {\bf #2}, #3 (#4)}
\def\NPA{{\em Nucl. Phys.} A}
\def\PRep{\em Phys. Rep.}
\def\PRC{{\em Phys. Rev.} C}
\def\FBSS{\em Few-Body Syst. Suppl. }

\begin{document}

\title{DEUTERON COMPTON SCATTERING BELOW PION PHOTOPRODUCTION THRESHOLD}

\author{ M. I. LEVCHUK}
\address{
B.I. Stepanov Institute of Physics, Belarusian Academy of
Sciences,\\ F. Skaryna prospect 70, 220072 Minsk, BELARUS
\\E-mail: levchuk@dragon.bas-net.by}

\author{A. I. L'VOV}
\address{
P.N.  Lebedev Physical Institute, Russian Academy of
Sciences,\\ Leninsky prospect 53, 117924 Moscow, RUSSIA
\\E-mail: lvov@x4u.lpi.ruhep.ru}

\maketitle

\abstracts
{
Deuteron Compton scattering below pion photoproduction threshold is
considered in the framework of the non-relativistic diagrammatic
approach with the Bonn OBE potential.  The complete gauge-invariant
set of diagrams is taken into account which includes resonance
diagrams without and with rescattering and diagrams with one- and
two-body seagulls.  The obtained results are compared with
predictions of other models and with experimental data. A
possibility of determining isospin-averaged electromagnetic
polarizabilities of the nucleon is discussed.
}

%%%%%%

The electric $\alpha_N$ and magnetic $\beta_N$ polarizabilities of the
nucleon are struc\-ture parameters characterizing the ability of the
nucleon to be deformed in external electromagnetic fields.  In the case
of the proton, the polarizabilities have been successfully measured in
low-energy $\gamma p$-scattering.  The polarizabiliti\-es of the
neutron have so far been measured only in low-energy experiments on
neutron transmission by the lead and on quasi-free Compton scattering
off deuterons, both giving not very certain results.  Elastic $\gamma
d$-scattering provides another attractive option for measurements. In
the present work we consider the amplitude of this process in the
framework of the non-relativistic diagram\-matic approach which
consistently takes into account electromagnetic interacti\-ons of
nucleons and those mesons which determine $NN$ interaction in the
deuteron, as contained in the non-relativistic versions of the Bonn
OBE poten\-tial \cite{Mach87,Mach89}.

In such an approach, the total $\gamma d$-scattering amplitude consists
of the resonance and seagull parts. The resonance part is determined by
two deuteron photodisintegration amplitudes of $\gamma d\to pn$ and $pn
\to \gamma' d'$, which are taken from Ref.~\cite{Lev95} with the
relativistic spin-orbit correction included, and by the full $T$-matrix
of rescattering the intermediate off-shell nucleons.  The seagull
operator, which involves both photons together, consists of one-body
parts (they are the Thomson term, a term with the nucleon
polarizabilities, and a relativistic spin-orbit contribution) and
two-body pieces which are determined by $\pi$, $\rho$, $\omega$,
$\sigma$, and $\delta$-meson exchanges consistently with the
$NN$-potential used.  Moreover, effects of the meson-nucleon form
factors, retardation effects in the meson propagators, and the
$\Delta$-isobar excitation are also included.

Our results are as follows.  For the success of determining the
isospin-averaged nucleon polarizabilities $\alpha = \frac12 (
\alpha_p+\alpha_n)$ and $\beta = \frac12(\beta_p+\beta_n)$ from $\gamma
d$-scattering, the one-body seagull contribution should be large at
energies of the major interest (50--100 MeV).  It is indeed the case,
as is shown in Fig.~1.  However, the resonance contribution and the
two-body seagull corrections are not small.  Only rescattering of
intermediate nucleons (which is most difficult for numerical
computations) has a little impact on the differential cross section
$d\sigma/d\Omega$.  Our results confirm findings of other approaches
\cite{Wey90,WWA95,LL95} that the rescattering decreases
$d\sigma/d\Omega$ by 7\% to 12\% at forward angles and energies 50 to
100 MeV and that it increases \cite{WWA95,LL95} $d\sigma/d\Omega$  by
7\% to 3\% at backward angles and the same energies.

\begin{figure}[hbt]
\psfig{figure=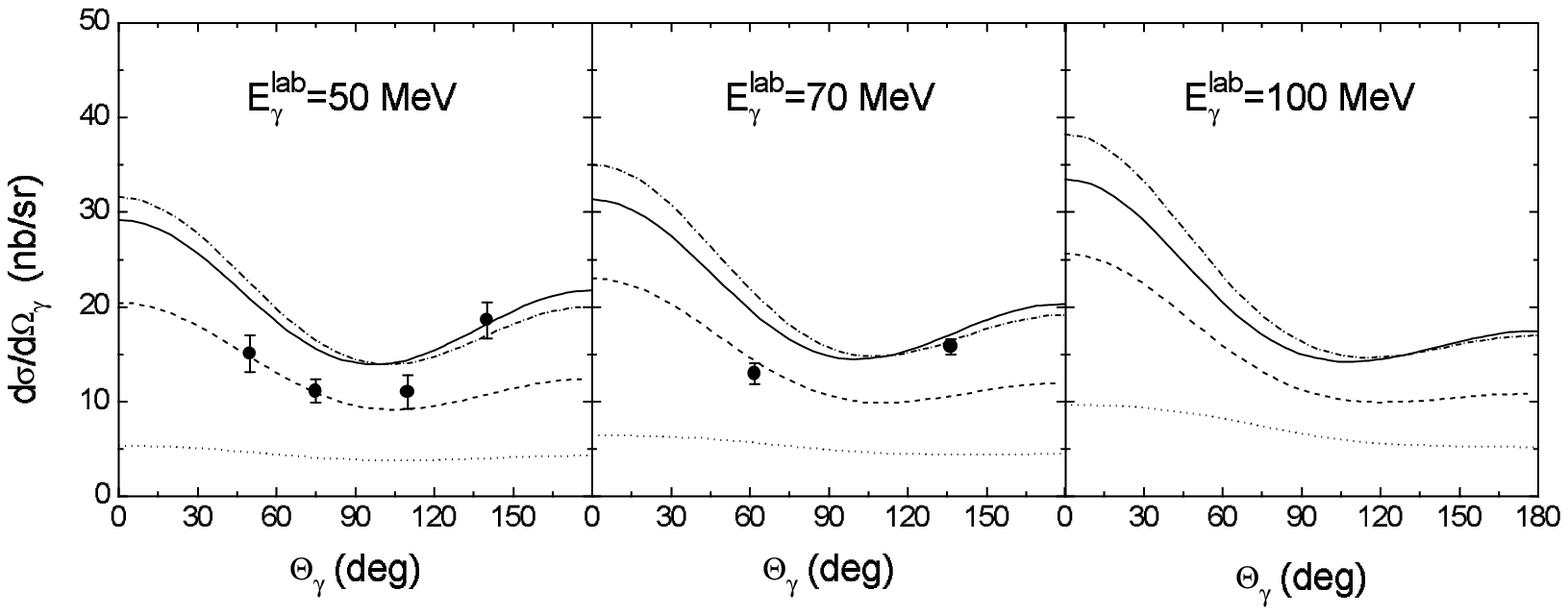,height=19cm}
\vspace{-14.8cm}
\caption{
Contributions of different diagrams to the differential CM-cross
section at 50, 70, and 100 MeV vs. the CM scattering angle
$\Theta_\gamma$.  Dotted lines: the resonance amplitude alone.
Dashed, dash-dotted, and solid lines: the one-body seagull, the
two-body seagull, and the rescattering contribution are successively
added.  For all curves $\alpha = \beta = 0$.  Data are from
Ref.~\protect\cite{Luc94}.}
\end{figure}

A comparison of our predictions with results of previous calculations
\cite{Wey90,WWA95,LL95,WA83} is shown in Fig.~2.  There is good
agreement with Refs.~\cite{Wey90,LL95} at 50 MeV and $\Theta_\gamma <
90^\circ$, but a big difference with predictions of Ref.~\cite{WWA95}
there.  Our previous calculation \cite{LL95} (``the minimal
gauge-invariant model") used a less sophisticated treatment of mesonic
contributions to electromagnetic currents and seagulls and did not
include the $\Delta$-isobar, retardation effects and the spin-orbit
correcti\-ons.  That is mainly why there is an increasing difference
between our present and older results when the energy increases.  
Much bigger difference is found at higher energies with 
Ref.~\cite{WWA95}, in which a very different angular dependence is 
obtained, and with Refs.~\cite{Wey90,WA83} at backward angles. We 
have no explanation for that. As a cross check of our results at 
$\Theta_\gamma=0^\circ$ and $\alpha=\beta=0$, we evaluated the 
Gell-Mann--Goldberger--Thirring dispersion relation for the 
spin-averaged forward amplitude of $\gamma d$-scattering using the 
available total cross sections of  deuteron photodisintegration and 
found very good agreement with the diagrammatic calculation which was 
better than 3\% for all energies below 100 MeV.

\begin{figure}[hbt]
\psfig{figure=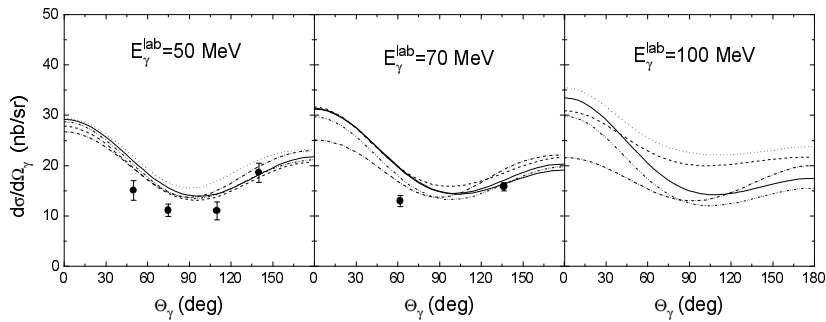,height=19cm}
\vspace{-14.8cm}
\caption{
Predictions of different models for the differential cross section at
three selected energies and $\alpha=\beta=0$.  Dashed lines:
Ref.~\protect\cite{Wey90}; dash-dotted lines:
Ref.~\protect\cite{WWA95}; dash-double-dotted lines:
Ref.~\protect\cite{LL95}; dotted lines: Ref.~\protect\cite{WA83}; full
lines: the present calculation.  Data are from
Ref.~\protect\cite{Luc94}.}
\end{figure}

When the nucleon polarizabilities are off, all the predictions
overshoot the available experimental data \cite{Luc94}.  With the
polarizabilities on, a much better agreement can be achieved.  Using
$\alpha + \beta=15$ (in units of $10^{-4}~ \rm fm^3$), as was estimated
\cite{LPS79} from dispersion relations, a few curves with different
$\alpha-\beta$ close to a theoretically expected range are shown in
Fig.~3.  The variation of $\alpha-\beta$ from 9 to 15 decreases the
backward differential cross section by 8\% and 20\% at 50 and 100 MeV,
respectively.  This supports a hope to determine $\alpha$ and $\beta$
and eventually the polarizabilities of the neutron from measuring
$\gamma d$-scattering.  Before, of course, reasons for large
disagreements between different theoretical computations must be
understood.

\begin{figure}[hbt]
\psfig{figure=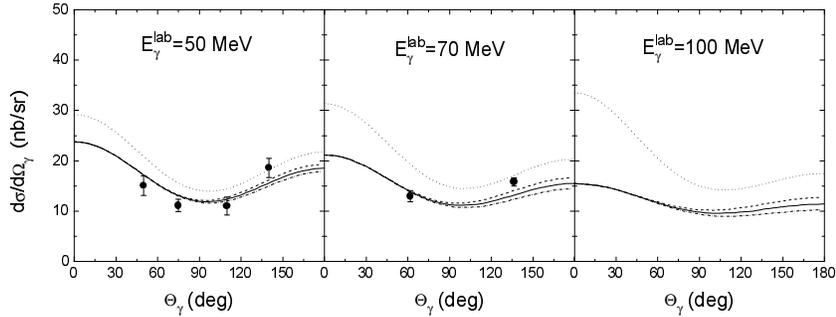,height=19cm}
\vspace{-14.8cm}
\caption{
The differential cross section at various isospin-averaged
polarizabilities.  Dashed, solid, and dash-dotted lines:
$\alpha-\beta=$9, 12, and 15, respectively, and $\alpha + \beta=15$
fixed (in units of $10^{-4}~\rm fm^3$). Dotted lines:  
$\alpha=\beta=0$.  Data are from Ref.~\protect\cite{Luc94}.}
\end{figure}

The presented calculation is able to describe all the experimental
points except those at the largest angle $\Theta_\gamma=140^\circ$.
Since at larger angles and hence at higher momentum transfers the
experimental separation of elastic $\gamma d$-scattering from inelastic
one ($\gamma d\to\gamma np$) is more difficult, it would be desirable
to independently confirm the strong increase of $d\sigma/d\Omega$ at
backward angles found in Ref.~\cite{Luc94}.  Currently, two experiments
at Lund \cite{Lund97} and \mbox{ Saskatoon \cite{SAL}} promise to
bring new data.

\bigskip
The work was supported by Advance Research Foundation of Belarus under
contract No. F96-155/503 and by the Russian Foundation for Basic
Research, grant 98-02-16534.


\section*{References}
\begin{thebibliography}{99}
\bibitem{Mach87}
R. Machleidt, K. Holinde, and Ch. Elster,
\Journal{\PRep}{149}{1}{1987}.

\bibitem{Mach89}
R. Machleidt, {\it Adv. Nucl. Phys.} {\bf 19}, 189 (1989).

\bibitem{Lev95}
M. I. Levchuk, {\it Few-Body Syst.} {\bf 19}, 77 (1995).

\bibitem{Luc94}
M. A. Lucas, PhD Thesis, Univ. Illinois at Urbana-Champaign, 1994.

\bibitem {Wey90}
M. Weyrauch, \Journal{\PRC}{41}{880}{1990}.

\bibitem {WWA95}
T. Wilbois, P. Wilhelm, and H. Arenh\mbox {\"o}vel,
\Journal{\FBSS}{9}{263}{1995}.

\bibitem {LL95}
M. I. Levchuk and A. I. L'vov, \Journal{\FBSS}{9}{439}{1995}.

\bibitem {WA83}
M. Weyrauch and H. Arenh\mbox {\"o}vel,
\Journal{\NPA}{408}{425}{1983}.


\bibitem{LPS79}
A. I. L'vov, V. A. Petrun'kin, and S. A. Startzev, {\it Sov. J. Nucl.
Phys.}  {\bf 29}, 651 (1979).

\bibitem{Lund97}
M. Lundin {\it et al.}, MAX-lab Activity Report 1996, p.237.
Edited by J. N. Andersen, R. Nyholm, M. Olofsson and S. L.
Sorensen.

\bibitem{SAL}
G. Feldman, private communication.

\end{thebibliography}
\end{document}